\documentclass[12pt]{article}
\pdfoutput=1 

\usepackage{amsmath,amssymb,amsfonts,epsfig,cite,setspace,bigstrut,framed}
\usepackage[all]{xy}
\usepackage{color}
\usepackage{pifont}
\usepackage{comment}

\makeatletter \@addtoreset{equation}{section} \makeatother
\renewcommand{\theequation}{\thesection.\arabic{equation}}

\begin{document}

\vskip 0.25in

\newcommand{\todo}[1]{{\bf\color{blue} !! #1 !!}\marginpar{\color{blue}$\Longleftarrow$}}
\newcommand{\nn}{\nonumber}
\newcommand\T{\rule{0pt}{2.6ex}}
\newcommand\B{\rule[-1.2ex]{0pt}{0pt}}

\newcommand{\cM}{{\cal M}}
\newcommand{\cW}{{\cal W}}
\newcommand{\cN}{{\cal N}}
\newcommand{\cH}{{\cal H}}
\newcommand{\cK}{{\cal K}}
\newcommand{\cT}{{\cal T}}
\newcommand{\cZ}{{\cal Z}}
\newcommand{\cO}{{\cal O}}
\newcommand{\cQ}{{\cal Q}}
\newcommand{\cB}{{\cal B}}
\newcommand{\cC}{{\cal C}}
\newcommand{\cD}{{\cal D}}
\newcommand{\cE}{{\cal E}}
\newcommand{\cF}{{\cal F}}
\newcommand{\cX}{{\cal X}}
\newcommand{\IA}{\mathbb{A}}
\newcommand{\IP}{\mathbb{P}}
\newcommand{\IQ}{\mathbb{Q}}
\newcommand{\IH}{\mathbb{H}}
\newcommand{\IR}{\mathbb{R}}
\newcommand{\IC}{\mathbb{C}}
\newcommand{\IF}{\mathbb{F}}
\newcommand{\IV}{\mathbb{V}}
\newcommand{\II}{\mathbb{I}}
\newcommand{\IZ}{\mathbb{Z}}
\newcommand{\re}{{\rm Re}}
\newcommand{\im}{{\rm Im}}
\newcommand{\tr}{\mathop{\rm Tr}}
\newcommand{\ch}{{\rm ch}}
\newcommand{\rk}{{\rm rk}}
\newcommand{\ext}{{\rm Ext}}
\newcommand{\bi}{\begin{itemize}}
\newcommand{\ei}{\end{itemize}}
\newcommand{\beq}{\begin{equation}}
\newcommand{\eeq}{\end{equation}}
\newcommand{\bea}{\begin{eqnarray}}
\newcommand{\eea}{\end{eqnarray}}
\newcommand{\ba}{\begin{array}}
\newcommand{\ea}{\end{array}}

\newcommand{\CA}{\mathbb A}
\newcommand{\CP}{\mathbb P}
\newcommand{\tmat}[1]{{\tiny \left(\begin{matrix} #1 \end{matrix}\right)}}
\newcommand{\mat}[1]{\left(\begin{matrix} #1 \end{matrix}\right)}
\newcommand{\diff}[2]{\frac{\partial #1}{\partial #2}}
\newcommand{\gen}[1]{\langle #1 \rangle}

\newtheorem{theorem}{\bf THEOREM}
\newtheorem{proposition}{\bf PROPOSITION}
\newtheorem{observation}{\bf OBSERVATION}

\def\theequation{\thesection.\arabic{equation}}
\newcommand{\setall}{
	\setcounter{equation}{0}
}
\renewcommand{\thefootnote}{\fnsymbol{footnote}}

\begin{titlepage}
\vfill
\begin{flushright}

\end{flushright}
\vfill
\begin{center}
{\Large\bf Scaling Behaviour of Quiver Quantum Mechanics}

\vskip 1cm

Heeyeon Kim\footnote{\tt heeyeon.kim@perimeterinstitute.ca}
\vskip 5mm
{\it Perimeter Institute for Theoretical Physics, \\31 Caroline Street North, Waterloo, N2L 2Y5, Ontario, Canada}

\end{center}
\vfill

\begin{abstract}
We explore vacuum degeneracy of Kronecker quiver with large
ranks, by computing Witten index of corresponding 1d gauged
linear sigma model.
For $(d-1,d)_k$ quivers
with the intersection number $k$, we actually counted index
of its mutation equivalent, $(d,(k-1)d+1)_k$, and find
exponentially large behaviour whenever $k\ge 3$. We close
with speculation on more general ranks of Kronecker quiver
including the nonprimitive cases.
\end{abstract}

\vfill
\end{titlepage}

\tableofcontents\newpage
\renewcommand{\thefootnote}{\#\arabic{footnote}}
\setcounter{footnote}{0}

\section{Introduction}

Wall-crossing phenomena and, more generally, the counting of
degeneracy of BPS states have played a central role during last
twenty years of development in superstring theory and supersymmetric
gauge theories. BPS states, which preserves part of supersymmetry of
the underlying field theory or gravity, are supposed to be robust
due to the short supermultiplet structure they come with. This
often allows us to probe otherwise inaccessible non-perturbative
aspects, such as U-dualty between different string theories or
even the grand idea of M-theory. With smaller supersymmetry preserved,
however, this is not always reliable, and a theory sometimes loses or
gains  BPS states suddenly under continuous deformation of
parameters, hence the wall-crossing phenomena. \cite{Seiberg:1994rs,Seiberg:1994aj,Ferrari:1996sv,Lee:1998nv}

Arguably, the most versatile framework where the intrinsic
degeneracy of BPS states can be counted is the quiver quantum
mechanics. This can be motivated rather simply by considering
collections of D-branes that are completely wrapped on supersymmetric
cycles in Calabi-Yau 3-fold. For compact Calabi-Yau's, such a
collection corresponds to BPS black holes while for noncompact ones,
to BPS states. This line of approach was initiated by Douglas et al., \cite{Douglas:2000ah,Douglas:2000qw}
and extensively developed by
Denef's pioneering work \cite{Denef} which inspired a lot of
physics and mathematics study of the wall-crossing. \cite{Denef:2007vg,KS,GMN1,GMN2,Gaiotto:2010be}

Nevertheless, direct and systematic counting of BPS states
starting with quiver quantum mechanics became available only
very recently. Problem of counting wall-crossing discontinuity
is often simpler than counting degeneracies themselves.
The so-called Coulomb approximation, which is
inspired by the multi-center picture of wall-crossing came
to fruition in Refs.\cite{deBoer:2008zn,Manschot:2010qz,Manschot:2011xc,Kim:2011sc,Sen:2011aa,Manschot:2013sya,Manschot:2014fua
}, which is however later seen to miss
a lot of physical states, when the quiver in question involves
oriented loops of bifundamental chiral multiplets. Honest
and comprehensive counting of Witten index was finally formulated
in Ref. \cite{Hori:2014tda,Cordova,Hwang} not only for quiver theories but also for general gauged
linear sigma models (GLSM), whose main result is phrased in terms of residue
formulae in the space of gauge field expectation values. Section 3
below will summarize the result.

Despite such developments, one aspects of the BPS state counting
remains much obscure. Namely, how the degeneracy scales as a
function of the increasing charge. This question itself is of
course very familiar in the context of BPS black holes, with
the obvious answer being "exponentially large," yet
most of definite answers on such a question are found in systems
where wall-crossing is either absent or rather simple. Black holes
that preserve four supercharges only and are BPS states
in the supergravity theory that preserves eight supercharges,
microstate counting in general is still unavailable. Isolating
the correct exponential behaviour from microstate counting based
on wrapped D-branes would be most useful, but this looks
rather difficult technically despite the above residue formula
as the number of residue integral to do grows linearly with
charge, and the number of poles to consider itself grows
exponentially fast.

On the other hand, the exponential behaviour of the degeneracy
turns out to be present also in certain field theory BPS states,
albeit to a lesser degree. This was initially anticipated by Kol \cite{Kol:1998zb}
on the basis of string-web picture of 1/4 BPS dyons in ${\cal N}=4$
field theory in four dimensions. While the former's argument
is somewhat anecdotal, a more definite counting using low energy
dynamics of solitons has shown how a highly charges states in
field theory can be equipped with an exponentially large degeneracy.
This exponential behaviour is different from that was anticipated with
black holes in the following sense. For BPS black holes in four
dimensional ${\cal N}=2$ supergravity, the scaling is such that
$$ \Omega_{BH}\sim e^{\#\Gamma^2}\ ,$$
for a large charge $\Gamma$, while for field theory BPS states the
anticipated scaling goes like
$$ \Omega\sim e^{\#\sqrt{\Gamma^2}}\ .$$
For either, one must deal with quiver quantum mechanics of
large rank $\sim \Gamma$, and the direct computation become
exponentially difficult even with the general formulation
of Witten index computation, unless one finds a mitigating
circumstance.

This note is an attempt to reproduce such an exponential behaviour
in the simplest possible nontrivial quiver, namely the Kronecker
quiver, in the hope of finding more general systematics of
large rank quiver dynamics. After a cursory introduction to
quiver quantum mechanics,  we go on
in section 3  to describe how the Witten index
of the Kronecker quiver of rank $d$ and $d-1$ can be computed via
diagrammatical representation of the residue formula,
following the method in Ref.~\cite{Hori:2014tda}.
A useful
middle step in this computation is a mutation which take
$(d-1,d)$ Kronecker quiver to $(d,(k-1)d+1)$ Kronecker quiver
where $k$ is the common intersection number. 
We briefly discuss how this duality for these quivers could be realized
in this context. Especially we obtained
\begin{equation}
\Omega_{(d,(k-1)d+1)_k} = \frac{1}{d}~[x^{(k-1)d+1}]g_k(x)
=\frac{k}{d((k-1)d+1)}{ (k-1)^2d +(k-1)\choose d-1}\ ,
\end{equation}
which implies the scaling bahaviour
\begin{equation}
\lim_{d\rightarrow \infty}\Omega_{(d-1,d)_k}=
\lim_{d\rightarrow \infty}\Omega_{(d,(k-1)d+1)_k} \sim e^{ f(k)d}\ ,
\end{equation}
where $f(k)=(k-1)^2\ln (k-1)^2-(k^2-2k)\ln (k^2-2k)$.
The result is consistent with the Euler number of quiver moduli spaces
which has been widely discussed in the mathematical literature \cite{Weist1,Weist2}
and in the context of spectral network \cite{Galakhov:2013oja}.
We close with some further speculations in section 5.

As this work was completed, a paper \cite{Cordova:2015qka} with some
overlap has appeared in the ArXiv.

\section{The Kronecker Quiver and Mutation}

The simplest setting where quiver quantum mechanics emerges is type IIB theory
compactified on Calabi-Yau 3-fold. The effective theory in the remaining
four dimensions carries ${\cal N}=2$ supersymmetry, and the BPS state thereof are
realized as D3-branes wrapped on special Lagrange subcycles of the Calabi-Yau.
When the cycle is rigid, as with $S^3$, the vector multiplet on the D3-brane
reduces to quantum mechanical vector multiplet whose content is the same
as ${\cal N}=1$ vector multiplet in four dimensions.
If D3 wraps the same cycle $d$ times, the gauge theory is elevated to
$U(d)$. For each such  D3-branes on $S^3$, we can associate a Fayet-Iliopoulos
constant $\zeta$ such that the low energy effective action carries a term
\begin{equation}
-\zeta \int dt \;{\rm tr}\, D\ ,
\end{equation}
where $D$ is auxiliary field in the vector multiplet. When a pair of such
wrapped D3's, say each wrapping $d_1$ and $d_2$ times, meets at $k$ intersections,
one find additional chiral multiplets in the bifundamental representation
$(\bar d_2, d_1)$. Such a quiver, which is the simplest possible nontrivial
class, is called Kronecker quiver. We have the supersymmetry constraint
\begin{equation}
d_1\zeta_1+d_2\zeta_2=0\ ,
\end{equation}
so we really have only one FI constant, say, $\xi\equiv d_1\zeta_1
=-d_2\zeta_2$. For negative $\xi$ the classical vacuum moduli space is
null and so is the Witten index. For positive $\xi$, the moduli space
is given as a Kaehler quotient,
\begin{equation}
\{ \Phi^{1,2,\dots,k}\in {\mathbb C}^{d_2}\times {\mathbb C}^{d_1}\;\vert\;
\Phi^\dagger \cdot\Phi =\zeta_1\}/S\left(U(d_1)\times U(d_2)\right)\ .
\end{equation}
The number of stable BPS bound states can be obtained by calculating indices of
these quiver quantum mechanics. 
Furthermore, the theory has $SU(2)_L\times U(1)_R$ R-symmetry, where $SU(2)_L$
came from the rotational symmetry of the three spatial direction. 
We can refine the indices by turning
on the fugacity ${\bf y}$ for $J_3+I$ of R-symmetry.
Due to the pioneering work of Reineke \cite{Reineke} and Manschot et al. \cite{Manschot:2010qz,Manschot:2011xc},
the systematic procedure of evaluating
the equivariant indices of mutually co-prime quiver $(d_1,d_2)$
are viable. However, since they are given in terms of very particular sum of
partitions of charges, it is hard to examine the large $d$ behaviour
with those formula. 

Indices for $k=1,2$ Kronecker quivers are well-documented and also
easy to compute directly. Note that the complex dimension of the
classical Higgs moduli space when $\xi>0$ is
$$ {\rm dim}{\cal M}_{(d_1,d_2)_k}
=k\cdot d_1\cdot d_2 - (d_1^2+d_2^2-1)= 1+(k-2) \cdot d_1\cdot d_2 -(d_1-d_2)^2\ .$$
This counting is misleading for nonprimitive
cases such as $(d,d)$, since the noncompact classical
moduli spaces open up along the Coulomb direction.
Direct evaluation of the standard formula leads to
fractional quantities, but with more care the true integral
index can be computed.
With $k=1$, the only primitive case with nonnegative dimension, thus nonempty
moduli space, is $d_1=d_2=1$. This case generates the most basic wall-crossing
pattern, corresponding to the so-called pentagon-identity, and
the index is
\begin{equation}
\Omega_{(1,1)_1}=1\ .
\end{equation}

With $k=2$, there are two classes with nonempty moduli spaces. The first is
$d_1=d_2=1$ with ${\rm dim}{\cal M}_{(1,1)_2}=1$  and the second is
$|d_1-d_2|=1$  with ${\rm dim}{\cal M}_{(d,d\pm 1)_2}=0$. Indices for
these can be inferred to the well-known spectrum of $SU(2)$ Seiberg-Witten
theory, and also have been computed directly by several different methods.
The answers are of course,
\begin{eqnarray}
\Omega_{(1,1)_2}=-\frac{1}{{\bf y}}-{\bf y}\ ,
\end{eqnarray}
and
\begin{eqnarray}\label{hyper}
\Omega_{(d,d-1)_2}=\Omega_{(d-1,d)_2}=1\ .
\end{eqnarray}
Seiberg-Witten wall-crossing formula implies that there are no other
nontrivial indices in this class, meaning also that all nonprimitive
Kronecker quivers have null (integral) index for $k=1,2$.

One most effective way to obtain Eq.~(\ref{hyper}) is to employ
the mutation map, which is beautifully motivated and described
in physics and mathematics of wall-crossing.
\cite{KS,Gaiotto:2010be,Alim:2011ae,Alim:2011kw}.
For general quivers with the nodes labeled with
charges $\gamma_i$ and intersection number $\langle\gamma_i,\gamma_j\rangle$
the left-mutation maps
\begin{equation}
\gamma_i\quad\rightarrow\quad  \left(\begin{array}{ccl}
-\gamma_k & \qquad & i=k \\ \\
\gamma_i+ [\langle\gamma_i,\gamma_k\rangle]_+ \gamma_k && \hbox{otherwise}
\end{array}\right.
\end{equation}
where $[a]_+$ is $a$ for positive $a$ and zero otherwise, while
the right mutation maps
\begin{equation}
\gamma_i\quad\rightarrow\quad  \left(\begin{array}{ccl}
-\gamma_k & \qquad & i=k \\ \\
\gamma_i+ [\langle\gamma_k,\gamma_i\rangle]_+ \gamma_k && \hbox{otherwise}
\end{array}\right.
\end{equation}
The basic assertion is that if we change $d_i$ as well to
keep the total $\sum_id_i\gamma_i$ fixed, the index remains
unchanged.

The mutation is not allowed for any nodes, however.
For wall-crossing quivers like this, there are the
questions concerning which node can be mutated and
which  of the two mutations should be taken, and the
answers to these questions are rather complicated.
Because Kronecker quivers have only two nodes and
because there is only one nonempty chamber, the
allowed mutation maps are uniquely fixed in that chamber as
\begin{eqnarray}
&&(d_1,d_2)_k \quad\Rightarrow \quad(d_2, k\cdot d_2- d_1)_k\cr
&& (d_1,d_2)_k \quad\Rightarrow \quad(k\cdot d_1-d_2, d_1)_k\ ,
\end{eqnarray}
under the left and right mutation respectively.
With $k=2$ Kronecker quivers, for instance, repeated usage
of this maps $(d,d-1)_2$ and $(d-1,d)_2$ quivers to $(1,0)_2$
and $(0,1)_2$, bringing us somewhat trivially back to Eq.~(\ref{hyper}).

In this note we are mostly concerned with $\Omega_{(d-1,d)_k}$ or
its mutation equivalent $\Omega_{(d,(k-1)d+1)_k}$, for $k\ge 3$.
It turns out that $k=3$ is a sort of watershed in that
the asymptotic scaling with $k\ge 3$ is qualitatively different
than those with $k< 3$. This may imply a similar scaling behaviour
for $\Omega_{(d,d)_k}$, which indicates that ${(d,d)_k}$ quivers have
nontrivial bound states for $k\ge 3$, unlike their counterpart
$k=1,2$, despite the flat Coulombic
directions opening up.

\section{Exact Formula for the Index of $(d-1,d)$ Kronecker Quiver}

\subsection{Index for the Kronecker Quivers}

The equivariant Witten index of interest is
\begin{equation}
\Omega_{\rm Q}={\rm tr} \left[(-1)^{2J_3} {\bf y}^{2J_3+2I}e^{-\beta H}\right]\ ,
\end{equation}
where we fixed the usual sign ambiguity of the index by choosing
$(-1)^F=(-1)^{2J_3}$. When we do this we should take care to remove the
center of mass part of the low energy dynamics, which is to say, to remove
one overall $U(1)$ decoupled from the rest of the dynamics.

The computation of this quantity for 1d gauged linear sigma model has been
extensively studied in \cite{Hori:2014tda,Cordova,Hwang,Ohta:2014ria}. 
For the Kronecker quiver $(d_1,d_2)_k$, the equivariant Witten index is computed as
Jeffrey-Kirwan residue \cite{JK,BV,SV},
\begin{eqnarray}\nonumber
\Omega_{(d_1,d_2)}(y) &=&\text{JK-Res}_\eta~\frac{(-1)^{k\cdot d_1\cdot d_2}}{d_1!\cdot d_2!}\left(\frac{1}{2\sinh z/2}\right)^{d_1+d_2-1}
\left(\prod_{p\neq q}^{d_1}\frac{\sinh[(x_p-x_q)/2]}{\sinh[(x_p-x_q-z)/2]}\right)
\\\label{mmm}
&&\times \left(\prod_{k\neq l}^{d_2}\frac{\sinh[(y_k-y_l)/2]}{\sinh[(y_k-y_l-z)/2]}\right)
\prod_{p,l,i}^{d_1,d_2,k}\left(\frac{\sinh[(x_p-y_l-a_{i}-z)/2]}{\sinh[(x_p-y_l-a_{i})/2]}\right)\ ,~~~~
\end{eqnarray}
where $\eta$ is chosen to be $\eta=(\zeta_1,\zeta_2)=(d_2,\cdots,d_2,-d_1,\cdots -d_1)$.
Here $x_i$'s and $y_j$'s denote the Cartans of $d_1$ (source) node and $d_2$
(sink) node respectively, and $a_i$'s are flavor fugacity for number of
arrows of the quiver. We also defined $e^{z/2}={\bf y}$.
In this section, we attempt to evaluate this expression
for various values of $d_1$, $d_2$ and $k$.

The residue formula above gets contribution from partial choice of
set of rank $r=d_1+d_2-1$ poles which are determined by the definition of the JK-residue.
At a singularity where exactly $r$ hyperplanes (each of which is defined
by a charge $Q_i\cdot u=0$) meet, the definition can be written as
\begin{eqnarray}\label{JK}
\hbox{JK-Res}_{\eta:\{Q_i\}}\frac{d^r u}{(Q_1\cdot u)(Q_2\cdot u)\cdots(Q_r\cdot u)}
= \left\{ \begin{array}{cc}\frac{1}{|{\rm Det} Q|} & \eta=\sum b_i^{>0} Q_i  \\ \\ \\
0 & {\rm otherwise}\end{array}\right\}\ ,
\end{eqnarray}
where we allowed constant shift of the pole location for
notational convenience. This procedure encounter some technical
difficulties when, at a contributing pole, more than $r$ such hyperplanes
meet resulting in the so-called degenerate cases. Constructive
procedures are known to deal with such cases, one of which
we will encounter in the Appendix. We concentrate on mutually co-prime
$d_1$ and $d_2$, where our choice of $\eta$ satisfies the regularity
condition where the definition of JK-residue for non-degenerate
point can be safely applied. \cite{Benini:2013xpa,JK,BV,SV}

In this note, we introduced all allowed chemical potentials,
$a$'s, to reduce degeneracies as much as possible.
Due to a powerful theorem \cite{Hori:2014tda}, dependence on these
chemical potential $a$ washes out whenever the classical moduli space is
compact, which is the case for Kronecker quivers with mutually
co-prime $d_{1,2}$. Degenerate singularities that cannot be resolved this way
will be separately addressed in the Appendix.

\subsubsection{$k=1$ and $k=2$}
First of all, we examine what this formula implies for the simplest case,
$k=1,2$. For $k=1$, since the classical moduli space is empty,
we expect $\Omega_{(d_1,d_2)}=0$ except for $\Omega_{(1,1)_1}(y)=1$.
The latter can be easily seen from a simple abelian residue integral.

For $d_1\neq d_2$, we can show that there is no charge set
which contributes to the JK-residue. First of all, in the Appendix,
we showed that the charge set which involves vector multiplet
can contribute only when $d_1 = d_2$.\footnote{
Although we have explicitly shown this for $k=1$ only,
we conjecture that for all values of $k$, when $d_1, d_2$ are mutually co-prime, 
singularities involving vector multiplet does not contribute to the integral.}
Hence, under the assumption that $d_1, d_2$ are mutually co-prime, the residue integral gets contribution only from poles of
chiral multiplets.
Suppose we have a set of chiral multiplet charges such that $\eta$ is in a positive cone of
the charges.
Then all the charges in this set should be connected to each other to meet this assumption,
and this implies that they collide at $x_i=y_k=0$ for all $i$'s and $k$'s.
This immediately produces $d_1(d_1-1)+d_2(d_2-1)$ additional zeros from the
vector multiplet numerator where additional degeneracy of the pole is $d_1d_2-(d_1+d_2-1)$.
Since $d_1(d_1-1)+d_2(d_2-1)>d_1d_2-(d_1+d_2-1)$ when $d_1,d_2\neq 1$,
the number of zeros always exceed that of additional poles.\footnote{Note that this counting is
equivalent to counting classical dimension of the quiver moduli space.}
Hence, chiral multiplets
cannot contribute either. Since there is no singularity which has non-vanishing residue
and meets the JK condition, we can say that $\Omega_{(d_1,d_2)_1}=0$ for all co-prime
$d_1$ and $d_2$ as expected.

Next, let us consider $(d_1,d_2)$ quivers with $k=2$, whose indices count the degeneracy of BPS spectra in
well-known $SU(2)$ Seiberg-Witten theory. First of all, the index of $(1,1)_2$ quiver
can be easily evaluated by adding two residue integrals of chiral multiplet poles,
\begin{equation}
\Omega_{(1,1)_2}= -\frac{1}{\bf y}-{\bf y}\ ,
\end{equation}
which correctly reproduces spin character of the four-dimensional BPS vector multiplet.
When $d_1=d_2$, as can be inferred from the argument in the Appendix,
residue integral gets contribution from
the vector multiplet, which again implies the existence of the flat direction in Coulomb branch.
For co-prime $d_1$ and $d_2$, when the index is expected to get
contribution only from chiral multiplets,
one can show that there are only two non-trivial cases, which are
$d_2=d_1+1$ and $d_1=d_2+1$.
Suppose that we turn off all the flavor fugacities. Since
all charges which contribute to the residue integral should be connected,
the rank $r$ singularity is located at $x_i=y_k=0$ for all $i$ and $k$.
At this point, the order of zeros minus the order
of additional poles are given by $d_1(d_1-1)+d_2(d_2-1)-[2d_1d_2-(d_1+d_2-1)]
=(d_1-d_2)^2-1$. Note that except $d_1= d_2+1$ and $d_2=d_1+1$,
the residue integral will vanish.

For $(d-1,d)_2$ (and equivalently $(d,d-1)_2$) quiver, we can further show that the
only non-trivial residue integral comes from a charge set where
each component $x_i$ of the first node appear exactly twice.
The contributing charge set is drawn in Figure \ref{fig:45} below.
\begin{figure}[h]
\centering
\includegraphics[width=0.4\textwidth]{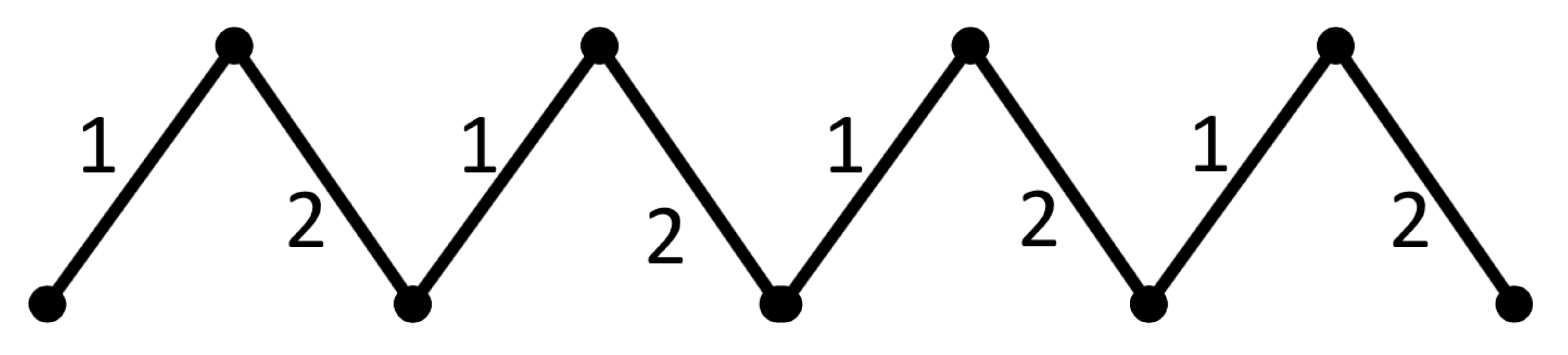}
\caption{Diagram of a contributing charge set for $(4,5)_2$ quiver.
This is only possible non-zero contribution, up to Weyl permutation.
The number denotes assignments of flavor fugacity.}
\label{fig:45}
\end{figure}
If there exist a component $x_i$ with more than three charges are connected,
one can see that the residue always vanishes or does not satisfy the JK-condition.
The residue integral at this fixed point can be readily evaluated.
If we set $y_1=0$, it follows that $x_n=na_1-(n-1)a_2,~(1 \leq n\leq d-1),~ y_m=(m-1)(a_1-a_2)
~(1 \leq n\leq d)$ are fixed locus of the above type.
Then, from the vector multiplets, we have a factor
\begin{equation}
\frac{1}{d!(d-1)!}\prod_{\substack{ n_1, n_2=1\\n_1\neq n_2}}^{d-1}\frac{\sinh[(n_1-n_2)(e_1-e_2)/2]}{\sinh[((n_1-n_2)(e_1-e_2)+z)/2]}
\prod_{\substack{m_1,m_2=1\\m_1\neq m_2}}^{d}\frac{\sinh[(m_1-m_2)(e_1-e_2)/2]}{\sinh[((m_1-m_2)(e_1-e_2)+z)/2]}\ ,
\end{equation}
and from the chiral multiplet, we have
\begin{equation}
\prod_{\substack{n,m=1\\n\neq m }}^{d-1,d}\frac{\sinh[((n-m)(e_1-e_2)-z)/2]}{\sinh[(n-m)(e_1-e_2)/2]}
\prod_{\substack{n,m=1\\n\neq m-1} }^{d-1,d}\frac{\sinh[((n-m+1)(e_1-e_2)-z)/2]}{\sinh[(n-m+1)(e_1-e_2)/2]}\ .
\end{equation}
Note that if we multiply two contributions, all the $\sinh$ factors cancel out each other.
Finally we should multiply by $d! (d-1)!$ which takes into account the contribution from
the Weyl equivalent fixed points.
We do not sum over $a_1\leftrightarrow a_2$, since it can be obtained from
left-right flipping of the diagram, which is a part of the Weyl group.
Hence, we correctly reproduce $\Omega_{(d-1,d)_2}(y)=1$ as expected.

\subsubsection{$k>2$}

Even though general recipe for calculating equivariant indices of the Kronecker quiver
is well-established, actual evaluation of the residue integral is very complicated
when degenerate singularities appear. However, almost all quivers with $k>2$
turns out to contain such singularity, as can be seen from the naive pole counting
after turning off the flavor fugacities.
If degenerate singularities exist, the answer depends crucially
on the order of taking each residue, which makes the evaluation for the large rank case
very involved.
See Appendix or \cite{Benini:2013xpa,JK,BV,SV} for how we should deal with of such singularities.

However, there are special types of two node quiver where we can explicitly show that
no degenerate singularity
appears even for $k>2$. These are the quivers of type $(d, (k-1)d+1)_k$. 
Interestingly, these quivers are known to be mutation dual to quivers of type
$(d-1,d)_k$, where large $d$ behaviour of the index has interesting physical
consequences \cite{Galakhov:2013oja}.

Especially, when the JK-residue gets contribution only from the non-degenerate poles,
the definition of the residue integral reduces to that of (\ref{JK}).
Now, suppose that we found a set of $d_1+d_2-1$ charges such that $\eta$ is
in a positive cone of the charge set.
Since they are all simple poles, evaluation of the residue corresponds to simple
substitution of these relations defined by $x_p=f_{p}(a_i)$ and $y_p=g_{p}(a_i)$,
where $f(a_i)$'s and $g(a_i)$'s are certain linear combinations of the flavor fugacities.
Then the contribution of this singularity to the residue integral is in the following form:
\begin{eqnarray}\nonumber
&&\prod_{p\neq q} \frac{\sinh[(f_p(a_i)-f_q(a_i))/2]}{\sinh[(f_p(a_i)-f_q(a_i)-z)/2]}
\prod_{k\neq l} \frac{\sinh[(g_k(a_i)-g_l(a_i))/2]}{\sinh[(g_k(a_i)-g_l(a_i)-z)/2]}\\
&&\times \prod_{f_p\neq g_k} \frac{\sinh[(f_p(a_i)-g_k(a_i)-z)/2]}{\sinh[(f_p(a_i)-g_k(a_i))/2]}\ .
\end{eqnarray}
Note that the pre-fector $\left(\frac{1}{2\sinh z/2}\right)^{d_1+d_2-1}$ are cancelled
by rank $d_1+d_2-1$ residue integral. Furthermore we multiplied $d_1!$ and $d_2!$, taking into account
the Weyl permutation of the solutions which give rise to the same contribution.
We expect that after we add up contributions from all singularities,
flavor fugacity cancels out \cite{Hori:2014tda} and end up with
a Laurent polynomial in $\bf y$.

Interestingly, the limit $z\rightarrow 0$ is well-defined for each of these term
and the residue becomes 1 at all such fixed points.
Hence, if we concentrate on the value of the Witten index only, the problem
reduces down to counting number of such set of charges which is in a positive
cone of corresponding $\eta$.
The following section is devoted to the evaluation of the Witten index for
$(d,(k-1)d+1)_k$ quiver, by counting all such contributing poles.
From now on, we denote set of \emph{nodes} by $i\in I$ and $j\in J$ which represent
each Cartans of source and sink of the Kronecker quiver respectively.
we will call \emph{contributing arrows} as a set of rank $d_1+d_2-1$
chiral multiplet charges $x_i-y_j-a_\alpha$
with specified flavor fugacities which passes JK condition and contributes non-zero value to the Witten index.

\subsection{Index and Large Rank Limit}
In this section, we focus on the explicit evaluation of the Witten index for
quivers of type $(d,(k-1)d+1)_k$, using the JK-residue formula derived
from the path integral. In the mathematical literature,
the Euler number of moduli space of these types of quiver was
extensively studied in \cite{Weist1,Weist2}. Our formula turns out to agree with
these results. \footnote{See also \cite{Cordova,Cordova:2015zra} which calculates the index of Kronecker
quiver by combining MPS formula and the JK-residue integral formula.}

The evaluation is done in the following procedure.
For quivers of type $(d,(k-1)d+1)_k$,
one can show following facts.
\begin{itemize}
\item[1.] Consider a set of contributing arrows. For each $i$, which labels
Cartan of the source, there are exactly $k$ set of arrows of type $x_i-y_j-a_\alpha$.

\item[2.] There is no degenerate contributing arrows.

\item[3.] Suppose that there exist two charges $x_i-y_j-a_\alpha$ and $x_i - y_j'-a_\beta$
with $a_\alpha= a_\beta$, for a source node $i$ of given contributing arrows.
Then these solution does not contribute to the integral. Similarly, for a sink node $j$,
if there exist two charges $x_i'-y_j-a_\alpha$ and $x_i - y_j-a_\beta$
with $a_\alpha= a_\beta$, this solution does note contribute to the integral.
\end{itemize}

These three statements are proven in the Appendix. Once we have these,
the procedure of finding all contributing arrows reduces down to that of
\cite{Weist1}. We can 
construct all sets of contributing arrows recursively, by \emph{gluing}
contributing arrows of the quiver with type $(1,k)_k$, which is defined as follow.

\begin{itemize}
\item[4.] \emph{Gluing} : Consider two sets of contributing arrows of type
$(d_1, d_2)_k$ and $(d'_1, d'_2)_k$ respectively. Let us denote the $I,J$ as
the set of source and sink nodes of the first quiver, and $I',J'$ as that of the latter quiver.
By \emph{gluing} we mean that identifying two sink node $j\in J$
and $j'\in J'$ with $j=j'=j_0$, in a way that it satisfies the condition 3.

\end{itemize}

Finally we also prove the following in the Appendix.

\begin{itemize}
\item[5.] All such configuration obtained from the gluing of $(1,k)_k$ type quiver
satisfies the JK condition.
\end{itemize}

Then, since we have shown that each non-degenerate contributing arrows contributes
1 to the Witten index, we can obtain index of the quiver just by counting number of
possible ways of gluing $d$ number of quivers of type $(1,k)_k$. 
We briefly review
the procedure below.

\begin{figure}[h]
\centering
\includegraphics[width=0.9\textwidth]{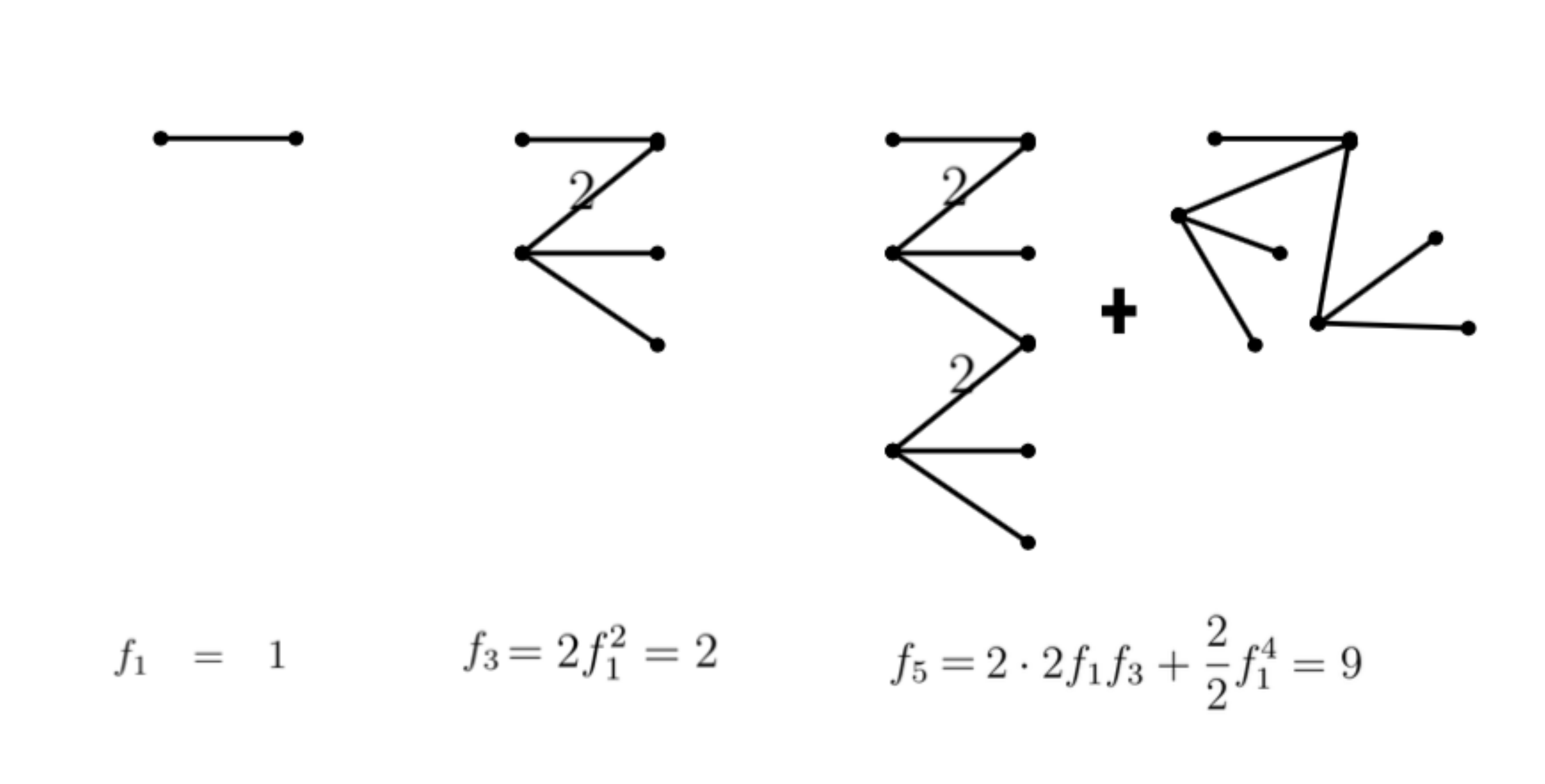}
\caption{Number of possible configuration of contributing arrows
which can be obtained by gluing several $(1,3)_3$
subquivers to $(1,1)_1$ type initial quiver. The numbers on the arrows denote
the choice of the flavor fugacities of corresponding arrows. Note that
it is crucial to divide by the order of symmetry group of the graph,
since it yields Weyl equivalent combinations, which we already have
taken into account in the calculation of the Witten index.}
\label{fig:MMM}
\end{figure}

Let us first illustrate the procedure with a simple example with $k=3$.
Suppose that we initially have $(1,1)_1$ type quiver with a flavor fugacity specified.
Then, by the condition 3, we can attach at most $k-1=2$ subquiver to this. 
By gluing $d$ copies of $(1,3)_3$ subquiver to this, we obtain
a quiver with $2d+1$ sink nodes.
Hence, if we denote $f_p$ as the number of possible configuration of such quiver
with $p$ sink nodes,
$f_{p}$ can be obtained from the expression of $f_k$ with $k<p$.
In the Figure \ref{fig:MMM}, this procedure for $d=0,1,2$ is drawn.
Note that it is crucial to divide by the order of symmtery group of the graph,
since it yields Weyl equivalent combinations, which we already have
taken into account in the calculation of the Witten index of each singularity.
One can easily find the expression for $f_{2d+1}$ in terms of  $f_k$'s with $k<2d+1$,

\begin{eqnarray}\nonumber
f_1&=&1\\\nonumber
f_2&=&0\\
f_3 &=&2f_1^2=2 \\\nonumber
f_4&=&0\\\nonumber
f_5 &=& 2\cdot 2 f_1f_3+\frac{2}{2}f_1^4= 9\\\nonumber
&\cdots
\end{eqnarray}

This procedure can be straightforwardly generalized to arbitrary
$k$ and $d$. If we denote $f_{n+1}^k$ by the number of contributing configuration
with $n+1$ sink nodes, it can be obtained via following relation.

\begin{eqnarray}\nonumber
f^k_{n+1} &=& ~{k-1\choose 1} \sum_{\substack{a_1,\cdots, a_{k-1}\\ \sum{a_i}=n}} f^k_{a_1}f^k_{a_2}\cdots f^k_{a_{k-1}}
+ {k-1\choose 2}\sum_{\substack{a_1,\cdots, a_{2(k-1)}\\ \sum{a_i}=n}} f^k_{a_1}f^k_{a_2}\cdots f^k_{a_{2(k-1)}}\\
\label{42}
&&+\cdots + {k-1\choose k-1}\sum_{\substack{a_1,\cdots, a_{(k-1)^2}\\ \sum{a_i}=n}} f^k_{a_1}f^k_{a_2}\cdots f^k_{a_{(k-1)^2}}\ ,
\end{eqnarray}
where the $i$-th term counts the number of ways of attaching $i$-many subquivers
with $n$ sink nodes in total, to the initial quiver of type $(1,1)_1$.
Each subquiver can be further decomposed into an arrow (which is attached to the initial quiver)
and $k-1$ sub-subquivers. When each of the latter contains 
$a_{p=1,\cdots k-1}$-many sink nodes respectively, it contributes $f_{a_p}$ factor to the $i$-th term.
Finally, the combinatorial factor $k-1 \choose i$ for $i$-th term takes into account the
number of ways of choosing flavor fugacity of arrows attached to the initial quiver, 
divided by order of the Weyl permutation of those.

Furthermore, this recursion relation can be neatly summarized  by introducing the
generating function
\begin{equation} \label{43}
f_k(x) = \sum_{n=1}^\infty f^k_n x^n\ ,
\end{equation}
with an auxiliary variable $x$. Then the recursion relation (\ref{42}) can be written
in terms of the algebraic equation which (\ref{43}) satisfies,
\begin{equation}
f_k(x) = x(1+f_k(x)^{(k-1)})^{(k-1)}\ .
\end{equation}
Finally, the actual generating function of the quivers in question, i.e., the one obtained by gluing $d$ subquivers
of type $(1,k)_k$ can be readily obtained in terms of $f_k(x)$ as
\begin{equation}
g_k(x) = f_k(x)^k\ ,
\end{equation}
which takes into account the fact that the initial quiver has been replaced by $(1,k)_k$ quiver,
yielding $k$ choices of initial gluing. Then $(k-1)d+1$-th coefficient of $g_k$
encode the number of contributing arrows of quiver type $(d,(k-1)d+1)_k$.

The coefficients of generating function which satisfies such algebraic
equation can be evaluated by the Lagrange inversion theorem.\cite{Weist1,Sedgewick,Drmota}
Especially, when a generating function satisfies an algebraic equation $f(x)=x(\phi(f(x))$
for a function $\phi(x)=1+ax^b$, we have \cite{Weist1}
\begin{equation}
[x^n](f(x)))^k =\frac{k}{n} {n \choose \frac{n-k}{b}} a^{(n-k)/b}\ ,
\end{equation}
where we denoted $[x^n]g(x)$ as $n$-th coefficient of a power series $g(x)$.
Using this theorem, we can extract the general expression for the Witten index,
\begin{equation}
\Omega_{(d,(k-1)d+1)_k} = \frac{1}{d}~[x^{(k-1)d+1}]g_k(x)
=\frac{k}{d((k-1)d+1)}{ (k-1)^2d +(k-1)\choose d-1}\ .
\end{equation}
Note that we further divided the answer by $d$ which corresponds to the choice of
the
initial quiver. The asymptotic behaviour of $d\rightarrow \infty$ for this
expression can be also calculated, \cite{Weist1} which we can write as
\begin{equation}
\lim_{d\rightarrow \infty} \frac{\ln \Omega_{(d,(k-1)d+1)_k}}{d} = (k-1)^2\ln (k-1)^2 - (k^2-2k) \ln (k^2-2k)\ .
\end{equation}
Recently, it was noted by Galakhov et al. \cite{Galakhov:2013oja} that
this exponentially large degeneracy of four-dimensional BPS states of conformal field theory
is something unexpected, since the naive dimensional analysis implies a bound
for the index,
$\log|\Omega(E)|\leq aV^{1/4}E^{3/4}$, where
$E$ is energy of state supported in a finite volume $V$. They showed that, by
carefully examining Denef's multi-center bound state formula \cite{Denef},
the radius of BPS bound state increases with the mass of the state.
This corrects the above bound by $\log|\Omega(E)|\leq a'E^{3/2}$,
which is consistent with the observed scaling behaviour of quiver quantum mechanics.

\subsection{Mutation Equivalences}

\begin{figure}[h]
\centering
\includegraphics[width=0.7\textwidth]{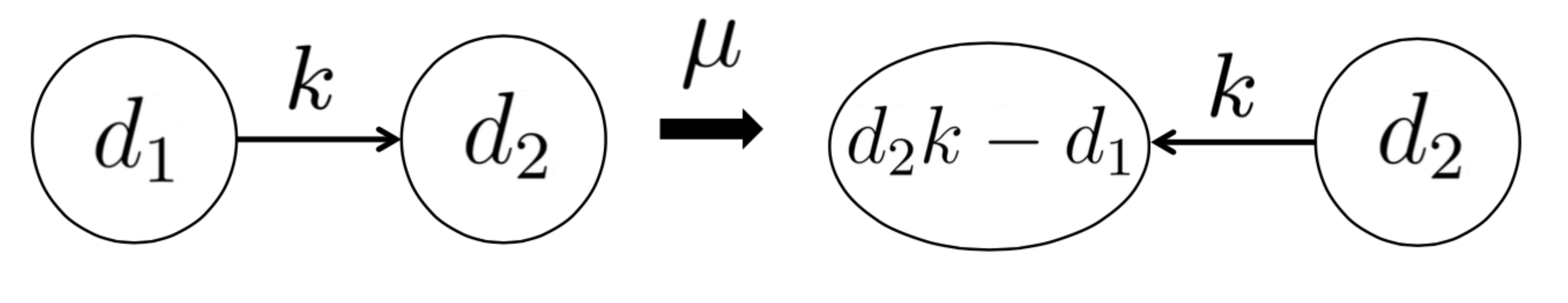}
\caption{Mutation equivalence for Kronecker quivers}
\label{fig:mu2}
\end{figure}

One of the most interesting properties of the 1d quiver quantum
mechanics is that they are expected to be invariant under certain
duality, which is called the \emph{mutation} equivalence.
In terms of the Kronecker quivers in question, it can be phrased
into the isomorphism between moduli space of $(d_1,d_2)_k$ quiver
with that of $(d_2k-d_1,d_2)_k$, as shown in the Figure \ref{fig:mu2}.

As a simple check, one can see the dimension of the classical moduli space of
these two quivers are the same.
Furthermore, we expect that, the equivariant indices of both
quivers are the same, i.e.,
\begin{equation}
\Omega_{(d_1,d_2)_k}(y) = \Omega_{(d_2,(k-1)d_1+1)_k}(y)\ ,
\end{equation}
which is calculated for region $\zeta_1>0$ and $\zeta'_1<0$ respectively. The index is
trivially same for the other side of the wall, since
both of them vanish.
In particular, we will examine how the relation
\begin{equation}\label{ggg}
\Omega_{(d,(k-1)d+1)_k}(y) = \Omega_{(d-1,d)_k}(y)
\end{equation}
can be realized, using the diagrammatic expression of the JK residue developed so far.

Before delving into general $d$ and $k$, we illustrate 
the proof of ($\ref{ggg}$) for one of the simplest pair
of the Kronecker quivers, $(2,3)_3$ and $(3,7)_3$ in detail,
where the contributing charge sets are explicitly written down.
In particular, we show that each fixed point determined by
a set of hyperplane charges can be mapped to each other in a particular way.

\begin{figure}[t]
\centering
\includegraphics[width=0.9\textwidth]{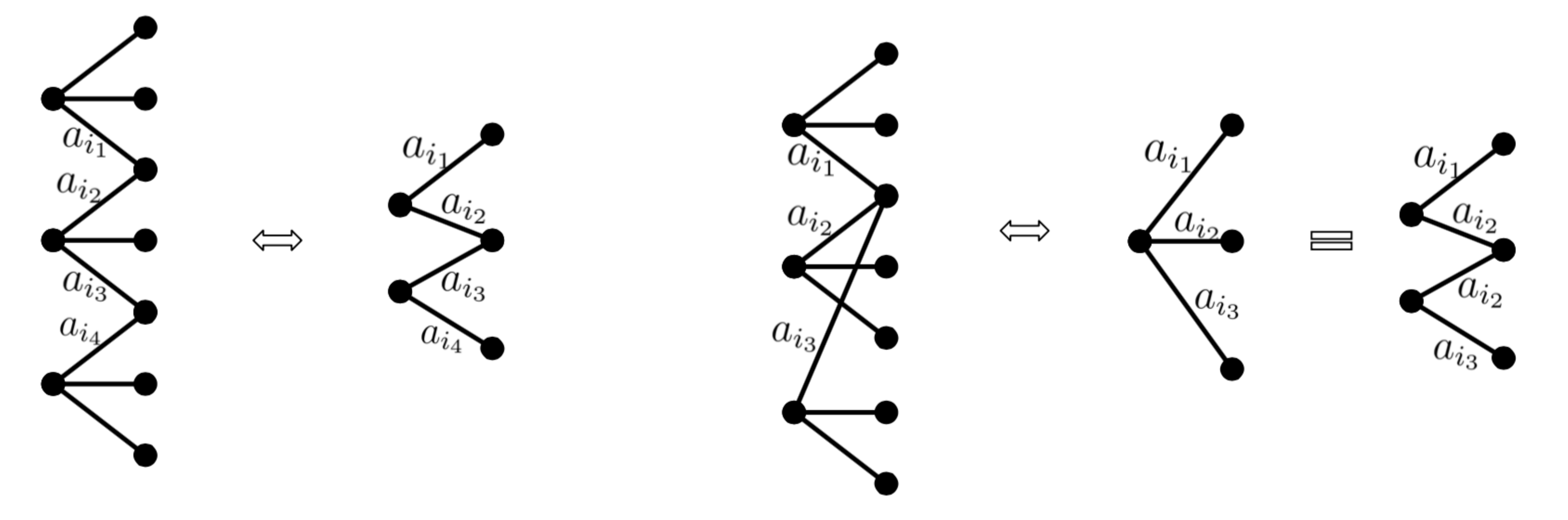}
\caption{Mapping of fixed points of two mutation equivalent quiver. Note that
each contributing arrows of $(d-1,d)_k$ appear as a subdiagram of 
contributing arrows of $((k-1)d+1,d)_k$ quiver, which has the same index.}
\label{hi}
\end{figure}

Let us denote the Cartans of $(2,3)$ quiver by
$u=(x_1,x_2,y_1,y_2,y_3)$. Then
the equivariant index of this quiver can be written
in the following expression
\begin{eqnarray}\nonumber
\Omega_{(2,3)}(y) &=&\text{JK-Res}_\eta~\frac{1}{2!}\frac{1}{3!}\left(\frac{1}{2\sinh[z/2]}\right)^4
\left(\prod_{p\neq q}^2\frac{\sinh[(x_p-x_q)/2]}{\sinh[(x_p-x_q-z)/2]}\right)
\\\label{bbb}
&&\times \left(\prod_{k\neq l}^3\frac{\sinh[(y_k-y_l)/2]}{\sinh[(y_k-y_l-z)/2]}\right)
\prod_{p,k,i}\left(\frac{\sinh[(x_p-y_k-a_{i}-z)/2]}{\sinh[(x_p-y_k-a_{i})/2]}\right)\ ,\qquad
\end{eqnarray}
where $a_i$'s are flavor fugacities for number of arrows.
Since they potentially carry degenerate singularities,
we need constructive definition of the JK-residue which can determine
the ordered set of contributing charges. (For the definition used here,
see \cite{SV} or section 2.4.3 of \cite{Benini:2013xpa}). For this purpose,
$\eta$ is chosen to be proportional to $\eta=(3,3,-2-\delta,-2,-2+\delta)$
with positive $\delta$.\footnote{
In order to safely apply the constructive definition of the JK-residue explained
in \cite{SV,Benini:2013xpa}, $\eta$
should not be at the boundary of flag defined by the sum of charges. Here,
$\eta$ should be
slightly shifted from $\zeta\sim (3,3,-2,-2,-2)$ since $\zeta$ can be spanned
by $\sum_i Q_i$, sum of all charge sets.}
If we carefully examine the JK-condition with this choice of $\eta$,
it turns out that the following ordered set of charges passes the JK condition,
and potentially non-zero.
\begin{eqnarray}\nonumber
&&x_1-y_1-a_{i_1}\\\nonumber
&&x_2-y_2-a_{i_2}\\\nonumber
&&x_2-y_3-a_{i_3}\\\label{fixed}
&&x_1-y_2-a_{i_4}\ ,
\end{eqnarray}

\noindent with their Weyl copies $x_1 \leftrightarrow x_2$ and all different
different set of flavor fugacities.
When $a_{i_1}\neq a_{i_4}$, $a_{i_2}\neq a_{i_3}$ and $a_{i_2}\neq a_{i_4}$,
this singularity is non-degenerate. On the other hand,
if $a_{i_2}= a_{i_4}$, this singularity becomes degenerate which has two
additional poles $x_1-y_3-a_{i_3}$, $x_2-y_1-a_{i_1}$ and order two 
zero from $x_1=x_2$.\footnote{For the latter case,
the order of $y_1,y_2,y_3$'s are fixed by choice of the sign of $\delta$.}
Other than these, the residue integral vanish due to the zero's of vector multiplet.
These two different classes of singularities are illustrated in the second and the fourth diagram of
Figure \ref{hi} respectively.

Meanwhile, the equivariant index of $(3,7)$ quiver with three arrows can be obtained from
gluing three $(1,3)_3$ quivers as was shown in the last section.
It turns out that there are two topologically distinguished contributing arrows which
can be obtained. These are also illustrated in the first and the third diagram of Figure \ref{hi}.
The non-equivariant index can be evaluated simply by counting number
of ways of assigning flavor fugacities to each arrows up to symmetries of diagram.
This turns out to be $-12$ and $-1$ respectively.

Interestingly, one can show that these two types of diagrams are mapped
respectively to the two different types of fixed points of the $(2,3)$ quiver.
For this purpose, we rewrite a term in the integral in (\ref{bbb}) evaluated at (\ref{fixed})
with $a_{i_1}=a_1, a_{i_2}=a_2, a_{i_3}=a_1, a_{i_4}=a_3$
as follow,
\begin{eqnarray}\nonumber
&&~\frac{1}{2!}\frac{1}{3!}\left(\frac{1}{2\sinh[z/2]}\right)^4
\oint_{y_1=y_2+a_{3}-a_{1}} \oint_{y_3=y_2+a_{2}-a_{1}}
\left(\prod_{k\neq l}^3\frac{\sinh[(y_k-y_l)/2]}{\sinh[(y_k-y_l-z)/2]}\right)\\
&&\oint_{x_2=y_2+a_{2}}\oint_{x_1=y_1+a_{1}}
\left(\prod_{p\neq q}^2\frac{\sinh[(x_p-x_q)/2]}{\sinh[(x_p-x_q-z)/2]}\right)
 \prod_{p,k,i}\left(\frac{\sinh[(x_p-y_k-a_{i}-z)/2]}{\sinh[(x_p-y_k-a_{i})/2]}\right). ~~\qquad
\end{eqnarray}
Note that,
the integral over all $x_i$'s in the last line can be evaluated to the simple expression as
\begin{equation}\label{qqq}
2!\cdot (2\sinh[z/2])^2\prod_{\substack{y_i,a_i\in A \\ y_j,a_j \in A'}}\frac{\sinh[(y_i-y_j +a_i-a_j-z)/2]}{\sinh[(y_i-y_j +a_i-a_j)/2]}\ .
\end{equation}
Here we defined that $A$ is a subset of all combinations $y_i+a_i$ which is integrated over in the
above integral (this case $y_1+a_1$ and $y_2+a_2$), and $A'$ is the complement of
$A$ in which case we have 7 elements. The factor $2!$ takes into account the Weyl
equivariant singular point.

\begin{figure}[t]
\centering
\includegraphics[width=0.7\textwidth]{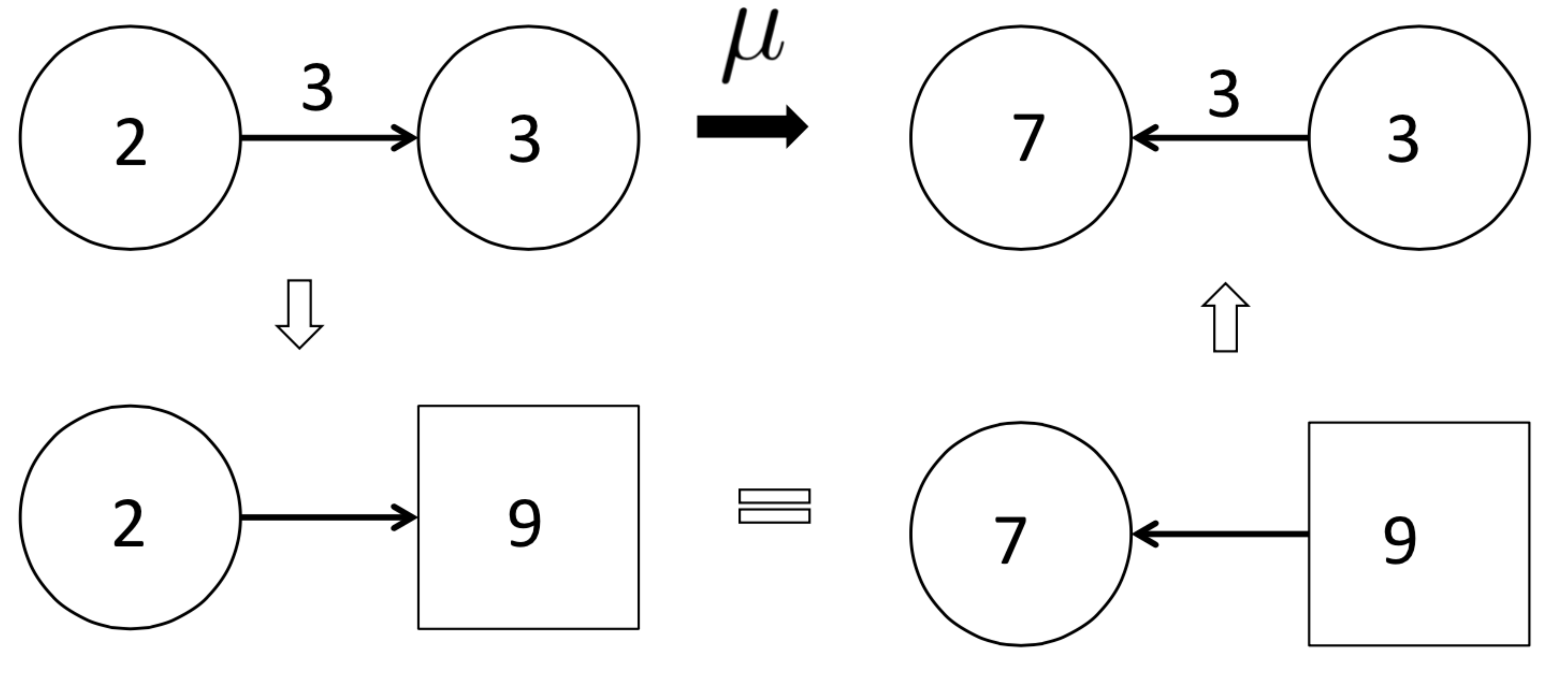}
\caption{Proof of mutation equivalence for $(3,7)$ and $(2,3)$ quiver with $k=3$}
\label{fig:mu3}
\end{figure}

On the other hand, if we look at $(3,7)$ quiver side, we have
\begin{eqnarray}\nonumber
\Omega_{(3,7)}(y) &=&-\text{JK-Res}_{\eta'}~\frac{1}{7!}\frac{1}{3!}\left(\frac{1}{2\sinh[z/2]}\right)^9
\left(\prod_{p\neq q}^7\frac{\sinh[(\tilde x_p-\tilde x_q)/2]}{\sinh[(\tilde x_p-\tilde x_q-z)/2]}\right)\\\label{mmq}
&&\times\left(\prod_{k\neq l}^3\frac{\sinh[(y_k-y_l)/2]}{\sinh[(y_k-y_l-z)/2]}\right)
 \prod_{p,k,i}\left(\frac{\sinh[(-\tilde x_p+y_k+a_{i}-z)/2]}{\sinh[(-\tilde x_p+y_k+a_{i})/2]}\right)\ .\qquad
\end{eqnarray}
Since all the poles are all non-degenerate, we can safely change
the order of integration without affecting the result. Suppose that we have contributing
arrows defined by the following assignments of flavor fugacities,
and fix the integration order as specified below.
\begin{eqnarray}\nonumber
-\tilde x_1+y_1+a_3\\\nonumber
-\tilde x_2+y_1+a_2\\\nonumber
-\tilde x_3+y_2+a_3\\\nonumber
-\tilde x_4+y_2+a_1\\\nonumber
-\tilde x_5+y_3+a_1\\\nonumber
-\tilde x_6+y_3+a_2\\\nonumber
-\tilde x_7+y_3+a_3\\\nonumber
-\tilde x_3+y_1+a_1\\\nonumber
-\tilde x_5+y_2+a_2
\end{eqnarray}
If we evaluate the residue integral for the first seven poles, it yields a factor
\begin{equation}
7!\cdot (\sinh[z/2])^7\prod_{\substack{y_i,a_i\in A \\ y_j,a_j \in A'}}\frac{\sinh[(y_i-y_j +a_i-a_j-z)/2]}{\sinh[(y_i-y_j +a_i-a_j)/2]}\ ,
\end{equation}
where the sets $A$ and $A'$ are exactly the same as what is defined in the expression (\ref{qqq}).
Again, we have a factor of $7!$ which takes into account different fixed points obtained by
the Weyl permutation of $\tilde x_i's$. In this way, for a given set of charges of $(2,3)$ 
quiver which satisfies the JK condition, 
we get unique rank 9 singularity for $(3,7)$ quiver after summing over the
Weyl permutations for both sides, and vice versa. Note that for charge sets mapped like this,
we are left with exactly the same integral over non-mutated nodes. 
This procedure is summarised in Figure \ref{fig:mu3}. Summing over all contributions, we end up with
\begin{equation}
\Omega_{(3,7)_3}(y)=\Omega_{(2,3)_3}(y)= -\frac{1}{{\bf y}^6}-\frac{1}{{\bf y}^4}-
\frac{3}{{\bf y}^2}-3-3{\bf y}^2 -{\bf y}^4 - {\bf y}^6\ ,
\end{equation}
which agrees with the formula obtained in various literatures. \cite{Reineke,Manschot:2011xc,Ohta:2014ria}
If we look at Figure \ref{hi} where we illustrated the mapping of each
contributing arrows, it is interesting to note that the contributing arrows of
$(2,3)_3$ quivers appear as a subdiagram of contributing arrows of
its mutation dual quiver. Once we pick all the arrows directly attached
to the glued nodes, it gives a
unique diagram which corresponds to the contributing arrows of
the dual quiver which has the same residue integral.

As can be seen in this example, 
once we classify all contributing arrows of both side of the dual pair,
this procedure may be generalised to arbitrary $d$ and $k$.
Suppose that, there exist an ordered charge set $\{Q_i\}$ which satisfies the
JK-condition for the $(d,(k-1)d+1)_k$ quiver.
Since this is non-degenerate, we can choose an order of the charge set 
in a way that we could integrate over
all $\tilde x_i$'s first, as was shown in the example.
Given this set, we uniquely pick a subquiver which includes all the
arrows directly connected to the glued node.
Out of this subquiver, we associate a unique bipartite quiver
as in Figure 4, with flavor fugacities assigned.
One can easily check that bipartite diagrams
are the only possibility which satisfies the JK-condition with $r=2d-2$ charges
for $(d-1,d)_k$ quiver.
Then, for non-degenerate poles of the latter, 
it is a direct generalization of the above example
that the residue of this pole is the same as that of
a pole of the original quiver.
For degenerate poles of $(d-1,d)_k$ quivers,
it needs to be separatley confirmed that the poles which can be
obtained this way saturate all set of degenerate poles
of the dual quiver, after shifting FI parameter properly
as in the above example.

\section{Comments}
To conclude, we would like to mention several interesting questions
which follows immediately.

\begin{itemize}
\item[1.] 
From the residue integral expression,
we can systematically obtain the index for the Kronecker quiver with non-coprime
dimension vector.
It would be interesting to study 
the scaling behaviour for non-coprime charges $\Omega((n,n)_k)$ directly.
As was partly shown in the Appendix, this type of quivers always
involve particular contributing charge sets which contain poles in the vector multiplet.
This implies that the index gets contribution from the flat direction
in the Coulomb branch, which can be inferred from the fact that
we generally get the fractional number for the poles of such type.

\item[2.] One of the most interesting problem remaining is to find a 
large-rank scaling behaviour of quivers with oriented loop, which is closely
related to counting the degeneracy of the single centered BPS black
hole in four-dimensional ${\cal N}=2$ theory.
It would be interesting to see if we can 
find any quiver with loop whose large-rank behaviour is accessible
through the residue integral formula.

\item[3.] The index formula described in this note
is given by particular sum over various residue integrals,
and the formula described by MPS \cite{Manschot:2010qz,Manschot:2011xc}
and Reineke \cite{Reineke} are also written in terms of
sum of various partitions. It would be interesting to see how 
our construction is related to the latter two expansion.
One of the pros of using the residue integral definition
is that it is rather easy to see the large rank behaviour and the mutation equivalence
of quivers, as shown in the example of the last section.
It remains to check if mutation equivalence holds for 
other types of the tree-like quiver and quiver with loops. 
\end{itemize}

\section*{Acknowledgement}
The author is grateful to Piljin Yi for fruitful comments and reviewing the manuscript,
and thanks to Seung-Joo Lee for useful discussion.
The research was supported by the Perimeter Institute for
Theoretical Physics. Research at Perimeter Institute is supported by
the Government of Canada through Industry Canada and by the Province
of Ontario through the Ministry of Economic Development and
Innovation. This research was made possible through the support of
a grant from John Templeton Foundation. The opinions expressed in
this publication are those of the author and do not necessarily reflect the
views of the John Templeton Foundation.

\appendix
\section{Technical Details}

\begin{itemize}

\item[\bf A1] \emph{For quivers of type $(d_1,d_2)_1$,
vector multiplet does not contribute when $d_1\neq d_2$.}

\emph{Proof.}
Suppose that there exist
a co-dimension $r$ singularity which contains poles from vector multiplet
$x_{i}-x_{j}=z$. Then since $\eta$ is taken to be positive for the first node,
it should be followed by a pole of type $x_{j}-x_{k}=z$ or $x_{j}-y_k=0$.
Since a charge set without any chiral multiplet cannot generate rank $r$
singularity, we need at least one pole from the chiral multiplet.
Thus in general, poles that contain vector multipets have a sequence
\begin{equation}\label{se}
x_{i_1}-x_{i_2}=z,~ x_{i_2}-x_{i_3}=z, \cdots, x_{i_k}-x_j=z,~ x_{j}-y_l=0
\end{equation}
as a part. However the last two relations
generate the equation $x_{i_k}-y_l-z=0$ which is zero of the hypermultiplet,
and the only possibility that a pole involving vector multiplet contributes
is that additional poles collide and become degenerate.
Let us scan all the charges possibly connected to this set. For $y_{p_1}\neq y_l$,
\begin{itemize}
\item[1)] $y_{l}-y_{p_1}=z$ : generates additional zero $x_j-y_{p_1}-z=0$.
\item[2)] $y_{p_1}-y_l=z$ : gives additional pole $x_{i_k}-y_{p_1}=0$.
\item[3)] $x_{i_0}-y_l=0$ : gives order two zero from $x_{i_0}-x_j=0$.
\item[4)] $x_{i_0}-x_{i_a}=\pm z$ : generates order two zero or go back to the same type.
\item[5)] $x_j-y_{p_1}=0$ : yields order two zero from $y_l-y_{p_1}=0$.
\item[6)] $x_{i_1}-y_{p_1}=0$ : No additional poles or zero.
\item[7)] $x_{i_a}-y_{p_1}=0$ : additional zero from $x_{i_{a-1}}-y_{p_1}=z$.
\end{itemize}
From these possibilities, we see that the sequences (\ref{se}) can contribute only
when it also contains a pole of type $y_{p_1}-y_l=z$.
However when $k>1$, introducing this pole yields another relation $x_{i_{k-1}}-y_{p_1}=z$
which generates a zero. In order to cancel this zero again, we further introduce
$y_{{p_2}}-y_{p_1}=z$ in a similar manner. This procedure continues to introduce
$k$ additional $y_p$'s, and we end up with the relations
\begin{eqnarray}\nonumber
&x_{i_1}-x_{i_2}=z,~ x_{i_2}-x_{i_3}=z, \cdots, x_{i_k}-x_j=z\ ,\\\nonumber
&y_{p_{k}}-y_{p_{k-1}}=z,~ y_{p_{k-1}}-y_{p_{k-2}}=z, \cdots, y_{p_1}-y_l=z\ ,\\\label{aaa}
&x_{j}=y_l,~ x_{i_k}=y_{p_1}, \cdots, x_{i_2}=y_{p_{k-1}},~ x_{i_1}=y_{p_{k}} ,
\end{eqnarray}
which involves exactly $k+1$ pairs of $x$'s and $y$'s.
If we try to add one more cartan $y_{k+1}$
which is related to the above set by adding
$x_{i_a}-y_{k+1}=0$ or $y_{p_{a}}-y_{k+1}=z$,
it always generates additional zeros of type $y_{p_{k+1-a}}-y_{k+1}=0$ or $x_{i_{k+1-a}}-y_{k+1}=z$. Finally, if we add $y_{k+2}-y_i=z$, this goes back to the argument of
the beginning, again introduces new $x_{i_{k+1}}$ which falls into type (\ref{aaa}).
Hence the set cannot be connected to other Cartans by adding additional
poles. However, when $d_1\neq d_2$, one can see that
the singularity involving this set cannot meet the JK condition.
This is simply because, sum of all components of any linear combination
$\sum_{i}a_iQ_i$ vanishes where $Q_i$'s are charge sets appearing in (\ref{aaa}).
The analysis implies that the vector multiplet cannot contributes for
co-prime $d_1$ and $d_2$, but they can contribute if $d_1=d_2$,
even for larger intersection number $k$.

\item[\textbf{A2}] \emph{Consider a quiver with type $(d, (k-1)d+1)_k$
and their particular set of $r=kd$ contributing arrows. In this set, for each $i\in I$,
there are exactly $k$ arrows of type $x_i-y_j-a_\alpha$.}

\emph{Proof.}
Note that the rank of this quiver is $kd$.
In order to satisfy the JK condition, there must exist a set of $kd$ positive coefficients
$a_{i,j,\alpha}$, such that
\begin{equation}\label{a1}
\zeta = \sum a_{i,j,\alpha}Q_{i,j,\alpha}= ((m-1)d+1,\cdots,(m-1)d+1,-d,\cdots, -d)
\end{equation}
holds. Here $Q_{i,j,\alpha}$ denotes a charge vector which gives a pole of type $x_i-y_j-a_\alpha$.
Now suppose that there exist $i\in I$ such that there are   
$m-q$ (for $0<q < m$) charges of a form $x_i-y_{j_p}-a_\alpha$.
Then there is at least one $j_{0}$ such that $a_{i,j_0,a}$, the coefficient for
charge of type $x_i-y_j-a_\alpha$, satisfies
\begin{equation}
a_{i,j_{0},\alpha} > \frac{(m-1)d+1}{m-q} > d.
\end{equation}
Then it is clear that existence of the coefficient of this property violates
$j_{0}$-th component of the equation (\ref{a1}).
Hence, for each $i$, there should be at least $k$ vectors of type $x_i-y_{j_p}-a_\alpha$.
Furthermore, since the number of
charges are exactly $kd$, we can say that
there should be exactly $k$ charges of type $x_i-y_{j_p}-a_\alpha$ for each $i$.

\item[\textbf{A3}] 1) \emph{For quivers with dimension vector $(d, (k-1)d+1)$, degenerate poles
do not contribute to the index.} 

2) \emph{Suppose that there exist two charges $x_i-y_j-a_\alpha$ and $x_i - y_j'-a_\beta$
with $a_\alpha= a_\beta$, for any $i$ in a contributing arrow set.
Then this set does not contribute to the integral. Similarly, for a sink node $j$,
if there exist two charges $x_i'-y_j-a_\alpha$ and $x_i - y_j-a_\beta$
with $a_\alpha= a_\beta$, this set does note contribute to the integral.
}

\emph{Proof.}
For non-degenerate singularities, poles of type 2) do not contribute
due to the zeros of the vector multiplets. 
However, these singularities may contribute for degenerate singularities
because of additional pole orders.
Note that, since the rank of the gauge group is $kd$ and there are only $k$ flavors,
degenerate singularity can appear only when one of the 2) is satisfied.

Let us briefly summarise how we can deal with such singularities in general.
Consider a gauge group with rank $r$. Then the JK residue gets contribution from
codimension $r$-singularities defined by
\begin{equation}
Q_i\cdot u =0\ ,
\end{equation}
for $i=1,\dots s$ with $s \geq r$. When $s=r$, definition of the JK residue integral
can be unambiguously written down as (\ref{JK}).
On the other hand, for $s>r$, which we call \emph{degenerate singularity},
the final answer depends very much on the order of taking residues,
and we need an alternative definition of JK residue which clarifies this ambiguity.
In the literature, there exist several equivalent definitions which could be
applied for these type of singularities. \cite{Benini:2013xpa,JK,BV,SV} Among them,
we are going to use the definition elaborated in \cite{BV,Hosomichi:2014rqa},
which we briefly summarize as below.

Suppose that we have rank $r$ degenerate singularity with charges $\{Q_{i=1,\cdots p}\}$
with $p\geq r$ colliding. From these, we fix the order of charges \emph{arbitrary},
and denote this ordered set as $\Delta = \{Q_{1},\cdots Q_{p}\}$.
Among them, there are $_{p}C_r$ choices of picking up
an ordered set with $r$-entry, which we denote as
$b=\{Q_{i_1},\cdots Q_{i_r}\}$. Finally, for each $b$, we assign so called
\emph{basic fraction},
\begin{equation}
\phi_b = \frac{1}{(Q_{i_1}\cdot u)\cdots (Q_{i_r}\cdot u)}\ .
\end{equation}
Out of these basic fraction $\phi_b$'s, we can build a set $b\in B$
such that $B$ form a rank $r$ basis of the basic fractions.
Then, from $B$, one can obtain any meromorphic function in $u$-plane
by taking derivatives and linear combinations of elements in $\phi_{b \in B}$.

Furthermore, the useful fact is that, for \emph{any} choice of
ordering of set $\Delta$, we can find a set $B$ such that we can define
iterated residue,
\begin{equation}\label{asd}
\text{Iterated-Res}_{b\in B} := \text{Res}_{Q_{i_r}\cdot u=0}\cdots \text{Res}_{Q_{i_1}\cdot u=0}
\end{equation}
with $b=\{Q_{i_1},\cdots Q_{i_r}\}$, which satisfy
\begin{equation}
\text{Iterated-Res}_{b\in B}~ \phi_{b'}=\delta_{bb'}\ .
\end{equation}
In particular, RHS of the definition (\ref{asd}) should be regarded as
iterated operation of the residue integral from the rightmost one, keeping
other components of $u$ at generic points.
With all these definitions, JK-residue for this degenerate point can be
simply written as
\begin{equation}
\text{JK-Res}(\eta, \{Q_1,\cdots, Q_{p}\}) = \sum_{b\in B,~\eta\in \text{Cone}(b)} \nu(b)  ~\text{Iterated-Res}_{b}\ ,
\end{equation}
where $\nu(b) = $~sgn(det $b$). 
With this alternative definition of the JK-residue, we are going to show that
for quiver with type $(d,(k-1)d+1)_k$,
there is no degenerate point which contributes to the integral.

As was mentioned above, for the Kronecker quivers of this type, degenerate
singularities from chiral multiplets can appear only when one of the following is satisfied.
\begin{itemize}
\item[1.] Contributing arrows contain two charges
$x_i-y_j-a_\alpha$ and $x_{i}-y_{j'}-a_\beta$ such that $a_\alpha=a_\beta$.
\item[2.] Contributing arrows contain two charges
$x_i-y_j-a_\alpha$ and $x_{i'}-y_{j}-a_\beta$ such that $a_\alpha=a_\beta$.
\end{itemize}
First, let us consider the first case. Suppose that there exist such $i\in I$ and $j,j'\in J$.
If both of $j,j'$ are not connected to any other source nodes, then this pole can be non-degenerate,
but they do not contribute since they generate the relation $y_j=y_{j'}$, which introduces
order 2 zero from the vector multiplet numerator.
If there exist charges of type $x_{i'}-y_j-a_\gamma$ or $x_{i''}-y_{j'}-a_\delta$ in this
 charge set, they generate additional relations $x_{i'}-y_{j'}-a_{\gamma}$
or $x_{i''}-y_{j}-a_\delta$. This implies that the corresponding pole additionally
collides at the singular point, so the singularity become degenerate.

Let us consider the second possibility. Suppose that we have such $i,i'\in I$
and $j\in J$ in a given set of contributing arrows.
It is rather easy to show that the second type of degenerate points never contribute
even if they generate degeneracy.
Suppose that there exist such $j\in J$ and $i,i'\in I$
which are assigned to a same flavor charge. Then from the argument of \textbf{A2},
we know that there exist at least $k-1$ $j'$ which is connected to $i$-th source node by
$x_{i}-y_{j'}-e_\gamma$, and $k-1$ $j''$  which is connected to $i'$-th source node
by $x_{i'}-y_{j''}-e_\delta$. Then they generate
$k-1$ additional relations of type $x_{i'}-y_{j'}-e_\gamma=0$, and also the
additional $k-1$ relations of type $x_{i}-y_{j''}-e_\delta=0$. In total, there
are $2(k-1)$ additional poles which additionally collide at this singular point.
However, they also generate at least $2(k-1)$
zeros of type $y_{j_a}=y_{j_b}$ for $a\neq b$ and two more from $x_i=x_{i'}$.
Since the order of additional zero $(2k)$ always
exceeds additional pole $(2(k-1))$, this degenerate singularity never contributes.

Finally, let us come back to the first degenerate singularity.
Suppose that $y_j=y_j'$ is the only relation which gives additional zero.
Note that, for this case, there
may exist a singularity where number of additional poles exceed or equal to the
number of additional zeros. However, one can show that they never satisfy the JK-condition.
First of all, since we have a freedom to choose the order of the charge set,
we set the order of charges $\Delta$ such that $x_i-y_j-a_\alpha$ $(:=Q_1)$ appear
at the very first position, and immediately followed by $x_i - y_{j'}-a_\alpha$ $(:=Q_2)$.
Then the ordered set $b$ contained in the set $B$ constructed as above procedure
should be one of the following.
\begin{itemize}
\item[1)] $\{Q_1,Q_2, \cdots\}$
\item[2)] $\{Q_1, Q_{i_1},\cdots\}$
\item[3)] $\{Q_2, Q_{i_2},\cdots\}$
\item[4)] $\{Q_{i_2},\cdots\}$
\end{itemize}
where $Q_{i_{1,2}}\neq Q_1 \text{ or } Q_2$. Note that the case $1)$ has
residue zero since this relation imposes $y_j=y_{j'}$ which gives arise the
order two zero. There are no additional poles at this point since
the other components of $u$ are kept generic, by the definition of the iterated residue.
Finally, consider the second, third and the fourth case.
If the set $b$ still has $k$ arrows connected to the source $x_i$, then it
defines rank $kd$ equation without any additional relations.
If this is the case,
the relation imposed by the original set $\Delta$ would have been overdetermined
and cannot define a singularity.
Hence, we can conclude that
there are less than $k$ arrows emerging from source $x_i$ in
the set $b$ of type 2),3) and 4). Going back to the proof of \textbf{A2},
we concluded that this type of charge set $b$ cannot satisfy the JK condition.
This procedure straightforwardly generalizes to the charge sets with larger number of additional zeros.

\item[\textbf{A4}] \emph{All configurations of contributing arrows obtained from gluing 
that of $(1,k)_k$ quiver satisfy the JK condition.}

\emph{Proof.}
This can be proven by induction.
First, let us assume that the statement is true for $(d, (k-1)d+1)_k$ type quiver.
It means that we have a set of $kd$ positive coefficients $a_{i,j,\alpha}$
such that the relation
\begin{equation}
\zeta_d = ((k-1)d+1,\cdots,(k-1)d+1,-d,\cdots, -d) = \sum a_{i,j,\alpha}
Q_{i,j,\alpha}
\end{equation}
holds, where $\{Q_{i,j,\alpha}\}$ is a charge set with $kd$ elements
which is obtained from the gluing procedure.
The assignment of flavor symmetry $\alpha$ obeys the
two conditions in 2) of \textbf{A3}.
Now, we attach a subquiver of type $(1,m)$ to this quiver
by identifying one of the sink node $y_j$ of the original quiver with that of new
subquiver $y_{j'}$ by $j=j'=j_0$.
We will see that for all such possible gluing, there exists
a positive coefficient set $\{a'_{i,j,\alpha}\}$ with $k(d+1)$ elements such that it satisfies
\begin{equation}\label{aa}
\zeta_{d+1} = ((k-1)(d+1)+1,\cdots,(k-1)(d+1)+1,-d-1,\cdots, -d-1) = \sum_{i, j \in A'} a'_{i,j,\alpha}Q'_{i,j,\alpha}\ ,
\end{equation}
where $Q'_{i, j,\alpha}$ is a charge of the new quiver.
We can show that positive coefficient $a_{i, j,\alpha}'$ that satisfies
(\ref{aa}) can be easily constructed from $a_{i, j,\alpha}$ of the original quiver.
Here, we denoted $A$ as a set of all nodes of original quiver
of type $(d, (k-1)d+1)_k$,
and $A'$ as a set of all nodes of new quiver $(d+1,(k-1)(d+1)+1)_k$.
First of all,
$a'_{i, j,\alpha}$, with $i, j\in A'\backslash A$, (i.e., when $i=d+1$)
can be uniquely identified as $a'_{d+1, j\neq j_0,\alpha} = d+1$.
Then it follows that $a'_{d+1, j_0,\alpha} = 1$. Since all the arrows are connected,
we can define numbers $a'_{i, j,\alpha}$ for all $i, j \in A$ through the following procedure.
First of all, consider the $j_0$-th component of the above equation.
Since $a'_{i, j_0,\alpha}$ satisfies
$\sum_{i,} a'_{i, j_0,\alpha} = d+1$ and $a'_{d+1, j_0,\alpha} = 1$,
we can set $a'_{i, j_0,\alpha} = a_{i, j_0,\alpha}$
for all $i\neq d+1$.
Then, consider $a_{i, j,\alpha}$'s for the above $i$'s and
$j\neq j_0$. If we look at $i$-th components of the equation,
they imply $a'_{i, j',\alpha}=a_{i, j',\alpha}+1$
with $j'\neq j_0$.
Next, consider the $j'$-th component of the equation. Similarly,
this implies that $a'_{i', j'',\alpha} = a_{i', j'',\alpha} $
for all $j'' \neq j'$. We can continue this procedure until all
$a'$'s are specified.
Since there is no cycle and every nodes are connected,
this procedure can uniquely fix all the coefficients
$a'_{i, j,\alpha}$ which are by construction all positive.
Finally, since the statement obvious holds for $d=1$,
we can say that for all $d$, $(d, (k-1)d+1)_k$ type quiver
obtained from the gluing method satisfy the JK condition.

\end{itemize}

\end{document}